\documentclass[showpacs,prd]{revtex4}
\usepackage{graphicx}
\usepackage{dcolumn}
\usepackage{bm}
\usepackage{amsmath}

\newcommand{\beq}{\begin{equation}}
\newcommand{\eeq}{\end{equation}}
\newcommand\Omggw{{\Omega_{\rm gw}}}
\newcommand\OmgX{{\Omega_{\rm X}}}
\newcommand\Omgacc{{\Omega_{\rm acc}}}
\newcommand\OmgSMBH{{\Omega_{\rm SMBH}}}
\newcommand\OmgM{{\Omega_M}}
\newcommand\Omglam{{\Omega_\Lambda}}
\newcommand\Omgrad{{\Omega_{\rm rad}}}

\begin{document}

\title{Gravitational-wave background from compact objects embedded in
AGN accretion disks}

\author{G\"unter Sigl}
\affiliation{APC~\footnote{UMR 7164 (CNRS, Universit\'e Paris 7,
    CEA, Observatoire de Paris)} (AstroParticules et Cosmologie),
  11, place Marcelin Berthelot, F-75005 Paris, France and\\
Institut d'Astrophysique de Paris, 98bis Boulevard Arago,
  75014 Paris, France}

\author{Jeremy Schnittman}
\author{Alessandra Buonanno}
\affiliation{Physics Department, University of Maryland, College Park,  MD 20742}

\date{\today}
 
\begin{abstract}
We consider a model in which massive stars form in a self-gravitating
accretion disk around an active galactic nucleus (AGN). These stars
may evolve and collapse to form compact objects on a time scale
shorter than the accretion time, thus producing an important family of
sources for LISA. Assuming the compact object formation/inspiral rate
is proportional to the steady-state gas accretion rate, we use the
intrinsic hard X-ray AGN luminosity function to estimate expected
event rates and signal strengths. We find that these sources will
produce a continuous low-frequency $(\lesssim 1 \mbox{ mHz})$
background detectable by LISA if more than $\sim 1\%$ of the accreted
matter is in the form of compact objects. For compact
objects with masses $\gtrsim 10 M_\odot$, the last stages of the inspiral
events should be resolvable above a few mHz, with rates as high as a
few hundred per year. 
\end{abstract}


\maketitle

\section{Introduction and Motivation}

Some time in the near future, the launch and operation of the Laser
Interferometer Space Antenna (LISA)~\cite{lisa} will issue in a new age of
gravitational wave (GW) astronomy. One of the major challenges that will
face GW astronomers is the successful discrimination
between instrumental noise, stochastic GW backgrounds,
and individual, resolvable GW sources. For example, the population of
galactic white dwarf binaries in close orbits will provide a major
contribution to the LISA noise curve in the $0.1-1$ mHz band~\cite{bender97}. 
At the same time, this confusion ``noise'' can also be
treated as a signal, and its shape and amplitude will provide
important information about the distribution and properties of white
dwarf binaries in the galaxy (e.g.\ \cite{cornish01}).

It has recently been proposed that, in the self-gravitating outer
regions of accretion
disks of active galactic nuclei (AGN), massive stars could form and
evolve, eventually collapsing into compact objects and merging with
the central black hole \cite{goodman03,levin}. This final
inspiral stage would be an important source of GWs in the LISA
band. Depending on the specific properties of the accretion disk and
the embedded compact objects, this population could
contribute to the GW background and also
produce individual resolvable signals. It is therefore a matter
of theoretical and practical interest to understand the nature
of such a population.

In this paper we attempt to derive a relationship between the
observable electromagnetic (EM) emission and the predicted GW
emission from AGN out to cosmological distances. In particular, we use
the hard X-ray luminosity function of Ueda et al.~\cite{ueda03} to infer
the accretion history of supermassive black holes (SMBHs) out to
redshifts of $z\sim 3$. Then we assume a few simple scaling factors,
such as the average (Eddington-scaled) accretion rate and the
efficiency of converting accretion energy to X-rays, and derive the
time-averaged GW spectrum that might be seen by LISA. Another important
model parameter is the black hole spin, which will determine the
radiative efficiency for gas accretion as well as the GW efficiency
for inspirals. Based on relativistic MHD simulations of accretion disks
\cite{devilliers03,gammie04,mckinney04}, as well as some recent
observations \cite{wang06}, there is growing evidence that most AGN
black holes should be 
rapidly (but {\it not} maximally) spinning, with $a/M\simeq 0.9-0.95$.

Depending on the specific model parameters, we find this background
could be an important class of LISA sources, similar in strength and
event rates to extreme mass-ratio
inspirals (EMRIs) from captured compact objects \cite{barack04}. As in
those sources, here too it is a matter of preference as to whether the
steady-state background
should be thought of as signal or noise. But for higher masses
(perhaps as large as $m\simeq 10^5 M_\odot$ \cite{goodman03}, provided
such objects are not tidally disrupted before producing significant
GW power,
disk-embedded compact objects should produce individual, resolvable
inspiral events with high signal-to-noise over a wide band of
frequencies. 

The outline of the paper is as follows: in Sec.~\ref{Xray_lf} we
give a brief overview of notation, and describe the cosmological X-ray
luminosity distribution from AGN. In Sec.~\ref{GW_signal} we
discuss the parameterization of the GW signal, from a single inspiral
event to an integral of all sources over redshift and AGN
luminosity. We then present the main results, showing the predicted GW
power spectra for a range of model parameters and also for parameters
that may depend on SMBH mass and redshift. In
Sec.~\ref{duty_cycle} we estimate the expected event rates and duty
cycle for LISA, which will be used to help distinguish between
stochastic signals and resolvable ones. In 
Sec.~\ref{discussion} we discuss implications for the LISA mission as
well as future GW observatories.

\section{The hard X-ray luminosity function of Active Galactic
  Nuclei}\label{Xray_lf}
We begin with a short discussion of notation. The results
derived below include a number of dimensionless parameters, most of
which can take values between 0 and 1. We divide these parameters into
three general classes: efficiencies, fractions, and
densities. Efficiencies, denoted by $\eta$, are believed to be
determined by more basic physics, and typically have more stringent
lower- and upper-limits. Fractions, denoted by $f$, are more
model-dependent parameters and less-well known than the efficiency
parameters, and thus have a larger range of acceptable values. Lastly,
various cosmological density parameters $\Omega$ are given as
fractions of the critical density $\rho_c\equiv 3H_0^2/(8\pi G)$. They
are most likely constrained to very small
values relative to unity. A summary of all the dimensionless
parameters used in the paper appears in Table \ref{glossary}, along
with acceptable and preferred values.

The rest mass in supermassive black holes is estimated as $\OmgSMBH
\simeq 2\times 10^{-6}$, determined from galaxy surveys using the
scaling relation between velocity dispersion and SMBH mass
(the well-known ``$M-\sigma$'' relation; \cite{salucci99, ferrarese00, yu02}).
The energy density in the diffuse X-ray background between 2 keV
and 10 keV is estimated as $\OmgX
\simeq2.6\times10^{-9}$~\cite{fabian92}, to which AGN are 
believed to make a considerable contribution~\cite{ueda03}.
Now, assume that a fraction
$f_{\rm acc}\leq1$ of the SMBH rest mass density is due to accreted
gas of cosmic density $\Omega_{\rm acc}$ (as opposed to mass gained
through mergers), which releases electromagnetic radiation with an
efficiency of $\eta_{\rm em}\leq1$.
We further assume that a fraction $f_{\rm X}\leq1$ of the electromagnetic
output is in the form of X-rays in the range $2-10$ keV. This implies
\begin{equation}\label{constants}
  f_{\rm acc}\,\Omega_{\rm SMBH} \simeq (1-\eta_{\rm em})\,\Omega_{\rm acc}
\end{equation}
and
\begin{equation}
  \Omega_{\rm X} \simeq \left\langle\frac{1}{1+z}\right\rangle\,
  f_{\rm unobsc}\,\eta_{\rm X}\,\Omega_{\rm acc}\,,
\end{equation}
where $\eta_{\rm X}\equiv f_{\rm X}\eta_{\rm em}\leq1$, and the average inverse
redshift of the sources $\left\langle(1+z)^{-1}\right\rangle\simeq0.4$
is due to the redshifting of radiation energy. The parameter $f_{\rm unobsc}$ 
in the above equation is the fraction of the intrinsic $X-$ray
luminosity that escapes the AGN region without obscuration or reprocessing.
Since from observations $\Omega_{\rm X}/\Omega_{\rm
SMBH}\simeq1.3\times10^{-3}$, this gives the constraint
\begin{equation}
  \left\langle\frac{1}{1+z}\right\rangle\,
  \frac{f_{\rm unobsc}\,f_{\rm acc}\,\eta_{\rm X}}{1-\eta_{\rm em}}\simeq
  \frac{\Omega_{\rm X}}{\Omega_{\rm SMBH}}\simeq1.3\times10^{-3}\,.
  \label{constants2}
\end{equation}
Since, for standard accretion disk theory $\eta_{\rm em}\lesssim0.3$
\cite{thorne74}, this immediately implies
\begin{equation}\label{eta_X}
  f_{\rm unobsc}\,f_{\rm acc}\,f_{\rm X}\,\eta_{\rm em} \gtrsim
  2\times10^{-3} 
\end{equation}
and thus
\begin{equation}
  2\times10^{-3}\lesssim f_{\rm unobsc},f_{\rm acc},\eta_{\rm
  X},f_{\rm X},\eta_{\rm em}\leq1\,. 
\end{equation}
The {\it lower limit} only depends on the fact that AGN
contribute a considerable fraction of the observed X-ray
background.

\begin{table}
\caption{\label{glossary} Glossary of dimensionless parameters, with
allowable and preferred values}
\begin{tabular}{ccccp{8cm}}
\hline
\hline
symbol & min & max & preferred & description \\
\hline 
$f_{\rm acc}$     & 0   & 1   & 1    & fraction of SMBH mass due to
                                       accreted gas \\
$f_{\rm co}$      & 0   & 1   & 0.01 & fraction of accreted matter in
                                       form of compact objects \\
$f_{\rm X}$       & 0   & 1   & 0.03 & fraction of EM radiation in
                                       X-rays \\
$f_{\rm Edd}$     & 0   & $\gtrsim 1$ & 0.1 & typical fraction of 
				       Eddington luminosity/accretion rate \\
$f_{\rm unobsc}$  & 0   & 1   & 0.3  & fraction of emitted X-rays not
				       absorbed/reprocessed\\
\hline 
$\eta_{\rm em}$   & 0   & 1   & 0.2  & accretion efficiency of
                                       converting gas to EM radiation \\
$\eta_{\rm X}$    & 0   &$\eta_{\rm em}$& 0.006 & accretion efficiency of
				       converting gas to X-rays \\
$\eta_{\rm gw}$   & 0   & 1   & 0.2  & accretion efficiency of
				       converting compact objects 
				       to GW radiation \\
\hline 
$\Omgacc$         & 0   & $\OmgSMBH$ & $2\times 10^{-6}$  
                                     & fraction of critical density 
                                       in accreted gas \\
$\OmgX$           & 0   & $\Omgrad$  & $2.6\times 10^{-9}$ 
                                     & fraction of critical density 
				       in X-rays (2-10 keV) \\
$\OmgSMBH$        & 0   & $\Omega_M$ & $2\times 10^{-6}$ 
                                     & fraction of critical density 
				       in SMBHs \\
$\Omggw$          & 0   & $\OmgSMBH$ & $2\times 10^{-10}$ 
				     & fraction of critical density in GWs\\
$\OmgM$           & 0   & 1   & 0.3  & fraction of critical density in matter\\
$\Omgrad$         & 0   & 1   & $5\times 10^{-5}$ & fraction of
				       critical density in radiation \\
$\Omglam$         & 0   & 1   & 0.7  & fraction of critical density in 
				       vaccuum energy \\
\end{tabular}
\end{table}

A growing consensus has been forming that SMBHs grow almost
exclusively by accretion, determined by linking luminosity
distributions to the SMBH mass distribution, suggesting $f_{\rm acc}\simeq1$.
\cite{salucci99,fabian99,yu02,elvis02,cowie03,marconi04,merloni04}.
A corollary
of this assumption is that most AGN should be rapidly spinning, with
$\eta_{\rm em} \simeq 0.15-0.3$ for dimensionless spin parameters of
$a/M \simeq 0.9-0.998$ \cite{thorne74}. As mentioned in the
introduction, recent MHD simulations suggest an upper limit to the
spin parameter, due to magnetic torques that remove angular
momentum from the inner edge of the disk, thus preventing the accreted
matter from spinning up the black hole to maximal spin
\cite{devilliers03,gammie04,mckinney04}. Thus we set
the fiducial spin parameter at $a/M = 0.95$, corresponding to
$\eta_{\rm em} \simeq 0.2$.

For $f_{\rm unobsc}\simeq0.3$~\cite{ueda03}, Eq.~(\ref{constants2})
would then require an X-ray efficiency of accretion of
$\eta_{\rm X}\simeq7\times10^{-3}$.
This would require either that the total radiative efficiency
$\eta_{\rm em}$ would have to be of the same order, or that most of
the electromagnetic emission is emitted in other bands. While such low
radiative efficiencies are certainly possible (e.g., ADAF models of 
Ref.~\cite{narayan94}), during the periods of largest growth and thus highest
AGN activity, the accretion disks should be radiatively efficient.
Indeed, the X-ray fraction $f_{\rm X}$
is governed by bolometric corrections and is of the order of 0.03 if
AGN emission is dominated by infrared and optical
frequencies~\cite{fabian99,elvis02,marconi04,Miller:2006tw}.
This is consistent with the fact that the background energy
in the X-ray band from 2 to 10 keV is dominated by AGNs, whereas
the energy density in the infrared background which is about a factor
200 higher, is dominated by ordinary galaxies.

Pending a fuller understanding of the star formation mechanism in the
accretion disk, for now we simply assume that a certain fraction
$f_{\rm co}\leq1$ of the accreted
material is in the form of compact objects which will not get tidally
disrupted before plunging into the SMBH.  Since the astrophysical
parameters that actually determine this fraction are not well known,
we set it to a conservative value of $0.01$. If it were much higher,
the disk would be entirely fragmented and thus not efficiently emit EM
radiation. And as we will see below, a value much below $1\%$ would result
in a GW signal undetectable by LISA. We
further assume that a certain fraction $\eta_{\rm gw}$ of the rest
mass $m$ of these compact objects is emitted in GWs during the
inspiral event. Neglecting the plunge and ringdown stages (as well as
magnetic torques in the innermost disk), we will
generally set $\eta_{\rm gw}=\eta_{\rm em}$. 

Following the above notation, we now write the X-ray luminosity
$L_{\rm X}$ as a fraction $f_{\rm X}$
of the bolometric luminosity,  which in turn is a fraction $f_{\rm
  Edd}$ of the Eddington luminosity $L_{\rm Edd}$:
\begin{equation}\label{Lx}
  L_{\rm X}=f_{\rm X} f_{\rm Edd} L_{\rm Edd}(M) = f_{\rm X}
  \eta_{\rm em}\dot{M}_{\rm acc}c^2,
\end{equation}
where the gas accretion rate is $\dot{M}_{\rm acc}$.
The Eddington limit is a function only of the SMBH mass: $L_{\rm
Edd}(M)=1.3\times10^{38}(M/M_\odot)$ erg/s. Over the range of redshifts
and luminosities we are probing, typical accretion rates during the
period of maximum black hole growth are estimated
to be $f_{\rm Edd}\sim0.1$, but could likely be even greater than
unity \cite{begelman02,collin02,king03}. This rate is also derived
from a comparison of the luminosity distribution and the mass
distribution, via the efficiency parameter $\eta_{\rm em}$.
Note that Eq.~(\ref{Lx}), together with
$f_{\rm X}\sim 0.03$ and $f_{\rm Edd}\sim0.1$, 
implies that this luminosity function corresponds to SMBH
masses $10^5M_\odot\lesssim M\lesssim10^{10}M_\odot$. This is
consistent with typical SMBH masses inferred from velocity dispersion
observations. Note, however, that $f_{\rm Edd}$ varies during
the lifetime of an AGN and, strictly speaking, is a distribution for
each SMBH mass and redshift. In the present work, we will understand
$f_{\rm Edd}$ as an average during the active periods of AGNs, when
they actually contribute to the X-ray luminosity function used below.

Emission in the infrared and hard X-rays is less obscured
than in other bands and are thus more easy to observe. Since
infrared luminosity functions are more poorly known,
we will use the X-ray luminosity function to obtain a realistic
picture of AGN distributions.
We stress that all our luminosities are to be understood as 
{\it intrinsic}, i.e.\ as a measure of the X-rays produced directly by
the accreting gas, and before reprocessing and/or partial absorption
within the host galaxy \cite{ueda03,marconi04}. This is most natural
for our purposes as we are dealing
with electromagnetic and GW emission by accretion, so we are concerned
primarily with emission from the very inner-most regions of the
disk.

\begin{figure}[ht]
\includegraphics[width=0.55\textwidth,clip=true]{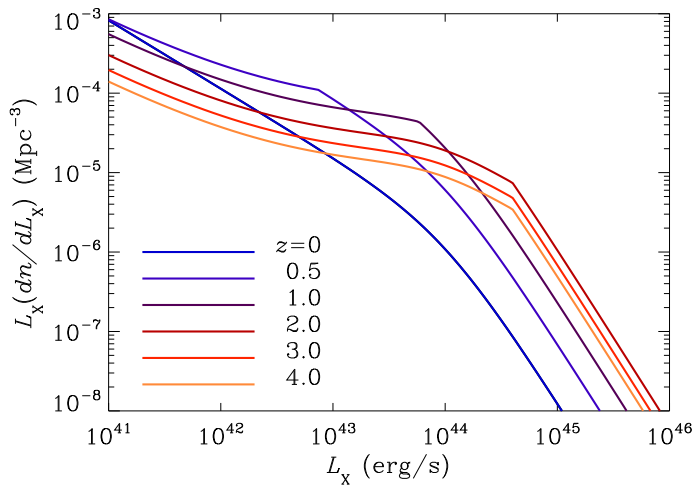}
\caption{\label{Lx_Ueda} Analytic fits to the hard X-ray luminosity
  function given in Ref.~\cite{ueda03}. Note that the average intrinsic
  luminosity (and thus AGN mass) increases with redshift.}
\end{figure}

The luminosity function is defined as the comoving number density $n$
of objects per logarithmic luminosity interval, with units of Mpc$^{-3}$.
We parametrize the intrinsic X-ray luminosity function per
comoving volume after Ref.~\cite{ueda03},
\begin{equation}\label{lumin1}
  L_{\rm X}\frac{dn}{dL_{\rm X}}(z,L_{\rm X}) = 
  \frac{A}{\left(\frac{L_{\rm X}}{L_*}\right)^{\gamma_1}+
  \left(\frac{L_{\rm X}}{L_*}\right)^{\gamma_2}}
  \times\left\{\begin{array}{ll}
  (1+z)^{p_1} & z<z_c(L_{\rm X})\,,\\
  \frac{(1+z)^{p_2}}{\left[1+z_c(L_{\rm X})\right]^{p_2-p_2}} & z\geq
  z_c(L_{\rm X})\,,
  \end{array}\right.
\end{equation}
for $L_{\rm min}\simeq10^{41.5}\,{\rm ergs}\,{\rm s}^{-1}\le L_{\rm X}\le
10^{46.5}\,{\rm ergs}\,{\rm s}^{-1}\simeq L_{\rm max}$. For a given
$L_{\rm X}$, the cut-off redshift is 
\begin{equation}\label{lumin2}
  z_c(L_{\rm X})=\left\{\begin{array}{ll}
  z_c^* & L_{\rm X}\geq L_a\\
  z_c^*\left(\frac{L_{\rm X}}{L_a}\right)^\beta & L_{\rm X}<L_a
  \end{array}\right.\,,
\end{equation}
and the best fit parameters in Eqs.~(\ref{lumin1}) and~(\ref{lumin2})
are given by $A=2.2\times10^{-6}\,{\rm Mpc}^{-3}$,
$L_*=10^{44}\,{\rm ergs}\,{\rm s}^{-1}$, $\gamma_1=0.86$,
$\gamma_2=2.23$, $p_1=4.23$, $p_2=-1.5$,
$z_c^*=1.9$, $L_a=10^{44.6}\,{\rm ergs}\,{\rm s}^{-1}$,
and $\beta=0.335$ (note that our value for the parameter $A$ differs
from Ref.~\cite{ueda03} by a factor of $(\ln 10)$ due to our slightly different
definitions of the luminosity function). The distribution
function defined by
Eq.\ (\ref{lumin1}) is plotted in Fig.~\ref{Lx_Ueda} for a range of redshifts. The ``anti-hierarchical
growth'' of AGN is evident from the increasing average luminosity with
redshift: at early times, {\it larger} black holes were growing
fastest, and most smaller black holes $(M \sim 10^6-10^7 M_\odot)$ are
thus relatively young \cite{marconi04,merloni04}.

With this luminosity function, we can calculate directly the diffuse
energy density in hard X-rays due to AGN:
\begin{equation}\label{Omega_X}
  \rho_c\Omega_{\rm X}\simeq f_{\rm unobsc}
  \int_0^\infty dz \left|\frac{dt}{dz}\right|\,\frac{1}{1+z}
  \int dL_{\rm X}\,L_{\rm X}\,\frac{dn}{dL_{\rm X}}.  
\end{equation}
In Eq.~(\ref{Omega_X}), cosmology enters through
$|dt/dz|=[(1+z)H(z)]^{-1}$ and, for a flat geometry,
\begin{equation}\label{cosmo}
  H(z)= H_0
  \left[\Omega_{\rm M}(1+z)^3+\Omega_{\Lambda}\right]^{1/2}\,.
\end{equation}
Throughout this paper we will assume a flat, $\Lambda$CDM Universe
with $\Omega_{\rm M}=0.3$, $\Omega_{\Lambda}=0.7$ (all other
contributions to the energy density are negligible), and $H_0=72~{\rm
  km}~{\rm s}^{-1}~{\rm Mpc}^{-1}$~\cite{kogut03}. Typically, we
will integrate the luminosity function out to redshifts $z_{\rm
  max}\simeq3$.

Reference ~\cite{ueda03} has demonstrated that, when integrated over
redshift and 
correcting for absorption and reprocessing, the luminosity function 
(\ref{lumin1}) explains practically all of the diffuse
extragalactic X-ray flux between $\simeq1\,$keV and a few hundred keV
measured by the {\it HEAO1}, {\it ASCA}, and {\it Chandra}
observatories. Combining Eqs.\ (\ref{lumin1}) and
Eq.~(\ref{Omega_X}), we obtain $\Omega_{\rm
  X}\simeq8.7\times10^{-9}f_{\rm unobsc}$, implying $f_{\rm unobsc}
\simeq0.3$. This is consistent with the estimate that about 60\% of
AGNs of luminosity $L_{\rm X}\lesssim10^{44}\,{\rm ergs}\,{\rm
  s}^{-1}$ are obscured ~\cite{marconi04}. The unobscured fraction
of these AGN dominates $\Omega_X$.

More information may come from future X-ray missions such as {\it
  Astro-E2/HXD, NeXT, Constellation-X,} and {\it XEUS}. With their
larger collecting area and greater sensitivity, these instruments will
allow us to extend the X-ray luminosity function to lower luminosities and
higher redshifts. Future IR observatories like the James Webb Space
Telescope will greatly improve our estimates for $f_{\rm X}$ and $f_{\rm
unobsc}$. 
There may also be significant AGN emission in soft $\gamma-$rays, as
recently reported from {\it INTEGRAL} observations \cite{beckmann06}. Future
$\gamma-$ray missions such as {\it GLAST} will further help us
understand the AGN energy budget at this end of the spectrum.

\section{The gravitational wave signal}\label{GW_signal}
\subsection{Individual inspiral spectrum}

We consider the inspiral of a single compact object of mass $m$ into a SMBH 
of mass $M\gg m$, motivated by scenarios such as those described in
the introduction \cite{goodman03, levin}. 
It is not known whether inspiraling black holes in accreting discs can 
acquire considerable eccentricity by interacting with the disc or 
with other orbiting masses, and if this eccentricity will have 
time to decrease due to GW radiation reaction as the small body 
spiral in. Considering the astrophysical uncertainties in deriving the
GW spectrum, we limit the analysis to the simplest inspiraling model,
i.e.\ the circular equatorial one. 

Using geometrized units such that $G=c=1$, a geodesic particle on an
equatorial, circular orbit around a Kerr 
black hole has an orbital frequency (as measured by an observer at
infinity) of \cite{shapiro83}
\begin{equation}
f_{\rm orb}(r) = \frac{\sqrt{M}}{2\pi(r^{3/2}\pm a\sqrt{M})}
\end{equation}
and specific energy
\begin{equation}
\frac{E(r)}{m} = \frac{r^2-2Mr\pm a\sqrt{Mr}}{r(r^2-3Mr\pm
2a\sqrt{Mr})^{1/2}}\,.\label{e_spec}
\end{equation}
We estimate the total energy emitted in gravitational waves down to a radius
$r$ as $E_{\rm gw}(r)=m-E(r)$.
The energy emitted in GWs between frequency $f$ and $f + df$ for such
an event is 
\begin{equation}\label{dEdf}
\frac{dE_{\rm gw}}{df} = \frac{dE_{\rm gw}}{dr}
\left(\frac{df}{dr}\right)^{-1}\,.
\end{equation}
We restrict the GW emission to the leading quadrupole formula, thus 
we consider only GW frequencies twice the orbital frequencies. 
In the Newtonian limit, we reproduce the well-known result 
(e.g.~\cite{thorne87,phinney01}) valid for circular orbits 
\begin{equation}\label{eps1}
  f\,\frac{dE_{\rm gw}}{df}(f) = \frac{m}{3}\,(\pi M f)^{2/3}.
\end{equation}
We will generally want to restrict Eq. (\ref{dEdf}) to a range of
frequencies $f_{\rm min} \le f \le f_{\rm max}$. Here,
$f_{\rm min}\equiv T_{\rm obs}^{-1}$ is the smallest resolution frequency-bin
determined by the mission mission lifetime $T_{\rm obs}$ and
$f_{\rm max}$ is the GW
frequency at the inner-most stable circular orbit (ISCO). The location
of the ISCO is given by the radius where $dE_{\rm gw}/df$ in
Eq.~(\ref{dEdf}) vanishes. It strongly depends on the spin of the black hole,
giving $f_{\rm max} \simeq 4-20$ mHz for a SMBH mass $10^6
M_\odot$ \cite{shapiro83}. Note that $\eta_{\rm gw}=
\int_0^{f_{\rm max}}df(dE_{\rm gw}/df)/m$ is given by Eq.~(\ref{dEdf}).

\begin{figure}[ht]
\includegraphics[width=0.55\textwidth,clip=true]{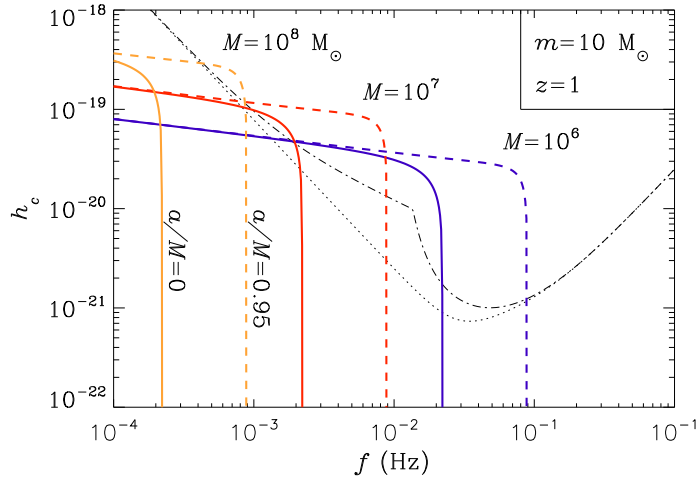}
\caption{\label{single_inspiral} Characteristic GW strain amplitudes
  for individual inspiral
  events, where a black hole with $m=10M_\odot$ merges with a SMBH of mass
  $M=10^6, 10^7, 10^8 M_\odot$ at a redshift of $z=1$. For each value
  of $M$, we show the spectra for two spin values, $a/M=0$ ({\it
  solid}) and $0.95$ ({\it dashed}). The {\it dot-dashed} and {\it
  dotted} lines are the sky-averaged LISA noise curves with and
  without the contributions from galactic binaries, respectively.}
\end{figure}

In Fig.~\ref{single_inspiral} we show the
GW spectrum for a single inspiral event for a range of black hole
masses and spins, plotting the so-called {\it characteristic} GW
amplitude $h_c(f)$ for single events, defined by
\begin{equation}\label{hc}
h_c^2(f) = \frac{2}{\pi^2}\,\frac{1}{r^2(z)}\,
\frac{d E_{\rm gw}}{df}(f_z)
\end{equation}
(compare with Figs.\ 3-7 in Ref.~\cite{finn00}), where $r(t)$ is the
comoving coordinate, $dr=(1+z)dt$.

\subsection{Integrated time-averaged spectrum}
To estimate the total contribution from all sources, we start by
integrating the GW energy density (per logarithmic frequency) over
redshift $z$, 
averaged over time scales large compared to all physical time
scales related to the sources. Following Ref.~\cite{phinney01}, we
have 
\begin{equation}\label{Omega1}
  \frac{d\rho_{\rm gw}(f)}{d \ln f} 
  =\int_0^\infty dz\,
  \frac{R(z)}{1+z}\left|\frac{dt}{dz}\right|
  f_z\frac{dE_{\rm gw}}{df_z}(f_z)\,,
\end{equation}
where $f_z\equiv(1+z)f$, and $R(z)$ is the rate of inspiral
events per comoving volume. Note the similarity to Eq.~(\ref{Omega_X}), 
the energy density in hard X-rays. 

To derive an expression for $R(z)$, we observe that the event rate for
a single AGN is simply the X-ray luminosity divided 
by the total X-ray energy emitted between inspiral events 
and thus for a given luminosity $L_{\rm X}$, the event rate is
\begin{equation}\label{R_LXz}
R(L_{\rm X},z) = \frac{dn}{dL_{\rm X}} \frac{L_{\rm X}}{E_{\rm X}}.
\end{equation}
Assuming equal efficiencies $\eta_{\rm gw}=\eta_{\rm em}$, this X-ray
energy can be written as
\begin{equation}\label{E_X}
E_{\rm X} = E_{\rm gw}\frac{f_{\rm X}}{f_{\rm co}}.
\end{equation}
Combining Eqs.~(\ref{R_LXz}), (\ref{E_X}) and integrating over the
luminosity distribution function, we get 
\begin{equation}\label{R_z}
  R(z) = \frac{f_{\rm co}}{f_{\rm X}} 
  \int dL_{\rm X}\frac{dn}{d\ln L_{\rm X}} \frac{1}{E_{\rm gw}}.
\end{equation}

Combining with Eq.~(\ref{Omega1}) and integrating over redshift, the total
(time-averaged) gravitational wave spectrum is 
\begin{equation}\label{Omega3}
  \frac{d \rho_{\rm gw}(f)}{d \ln f}= \frac{f_{\rm co}}{f_X}
  \int_0^\infty \left|\frac{dt}{dz}\right|\,\frac{dz}{1+z}
  \int_{L_{\rm min}}^{L_{\rm max}} dL_{\rm X}
  \frac{dn}{d \ln L_{\rm X}}
  \frac{1}{E_{\rm gw}} \frac{dE_{\rm gw}}{d\ln f_z}(f_z).
\end{equation}
The GW spectrum $E_{\rm gw}(f)$ from each individual AGN is a function
of the SMBH mass, which in turn is determined by the X-ray luminosity
through Eq.~(\ref{Lx}). Note that the integrated spectrum is
independent of $m$, as long as $m$ is small
enough so that the inspiral waveforms cannot be individually
resolved. One measure of this resolvability is the {\it duty cycle},
described in the next section.

Following Refs.~\cite{finn00,barack04}, we will want to compare directly
the stochastic background defined in Eq. (\ref{Omega3}) to the
spectral density of the detector noise $S_{\rm n}(f)$, which has units of
inverse frequency. In this case, $\sqrt{f S_{\rm n}(f)}$ will have units of
dimensionless strain. Averaging over the entire sky, weighted by the
LISA antenna pattern, gives
\begin{equation}\label{Srho}
S_h(f) = \frac{4}{\pi}\,\frac{1}{f^{3}}\,
\frac{d \rho_{\rm gw}^{\rm av}(f)}{d \ln f}.
\end{equation}
Throughout the paper we use the so-called sky and detector averaged
instrumental spectral density for LISA, given by~\cite{finn00,barack04}: 
\beq
S^{\rm instr}(f)=
\left(6.12\times 10^{-51}\,f^{-4}+1.06\times 10^{-40} +6.12\times
10^{-37}f^2\right)~{\rm Hz}^{-1}\,, 
\eeq
augmented by the white-dwarf galactic confusion noise
\beq
S_h^{\rm gal}(f)=
1.4\times 10^{-44}\left(\frac{f}{1~{\rm Hz}}\right)^{-7/3}~{\rm Hz}^{-1}\,,
\eeq
and the white dwarfs extra-galactic confusion noise
\beq
S_h^{\rm ex-gal}(f)=
2.8\times 10^{-46}\left(\frac{f}{1~{\rm Hz}}\right)^{-7/3}~{\rm Hz}^{-1}\,.
\eeq
Thus, the total (instrumental plus confusion) noise is
\beq
S_{\rm n}(f)={\rm min}\left\{
S^{\rm instr}(f)/{\rm exp}
\left(-\kappa T^{-1}_{\rm mission} dN/df\right),~
S_h^{\rm instr}(f)+S_h^{\rm gal}(f)
\right\}+S_h^{\rm ex-gal}(f)\,.
\label{Shtot}
\eeq
Here, $dN/df$ is the number density of galactic white-dwarf binaries
per unit gravitational-wave frequency given by
\beq
\frac{dN}{df}=2\times 10^{-3}~{\rm Hz}^{-1}\left(\frac{1~{\rm
    Hz}}{f}\right)^{11/3}\,. 
\eeq

\subsection{Dependence on model parameters}

\begin{figure}[ht]
\includegraphics[width=0.45\textwidth,clip=true]{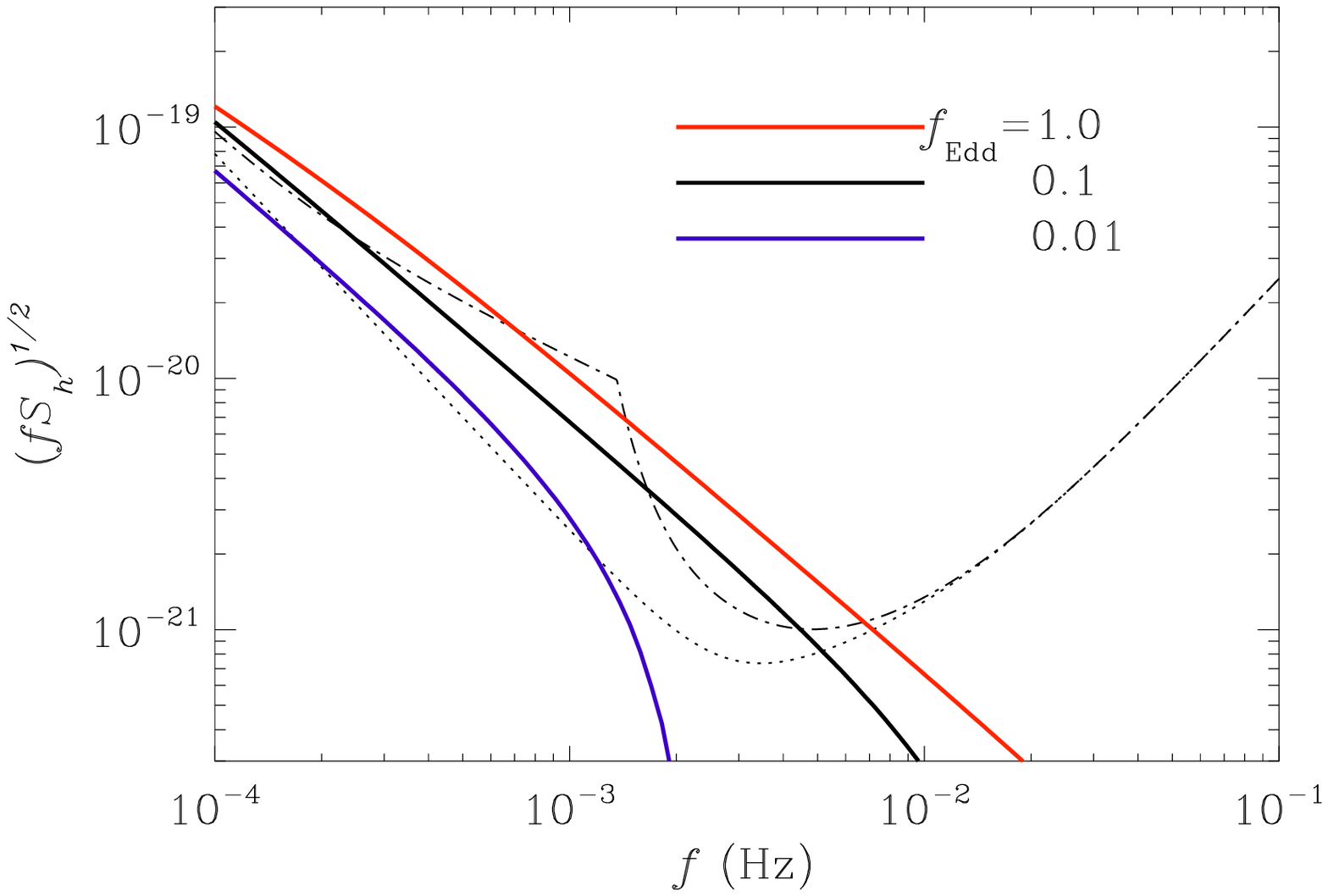}
\includegraphics[width=0.45\textwidth,clip=true]{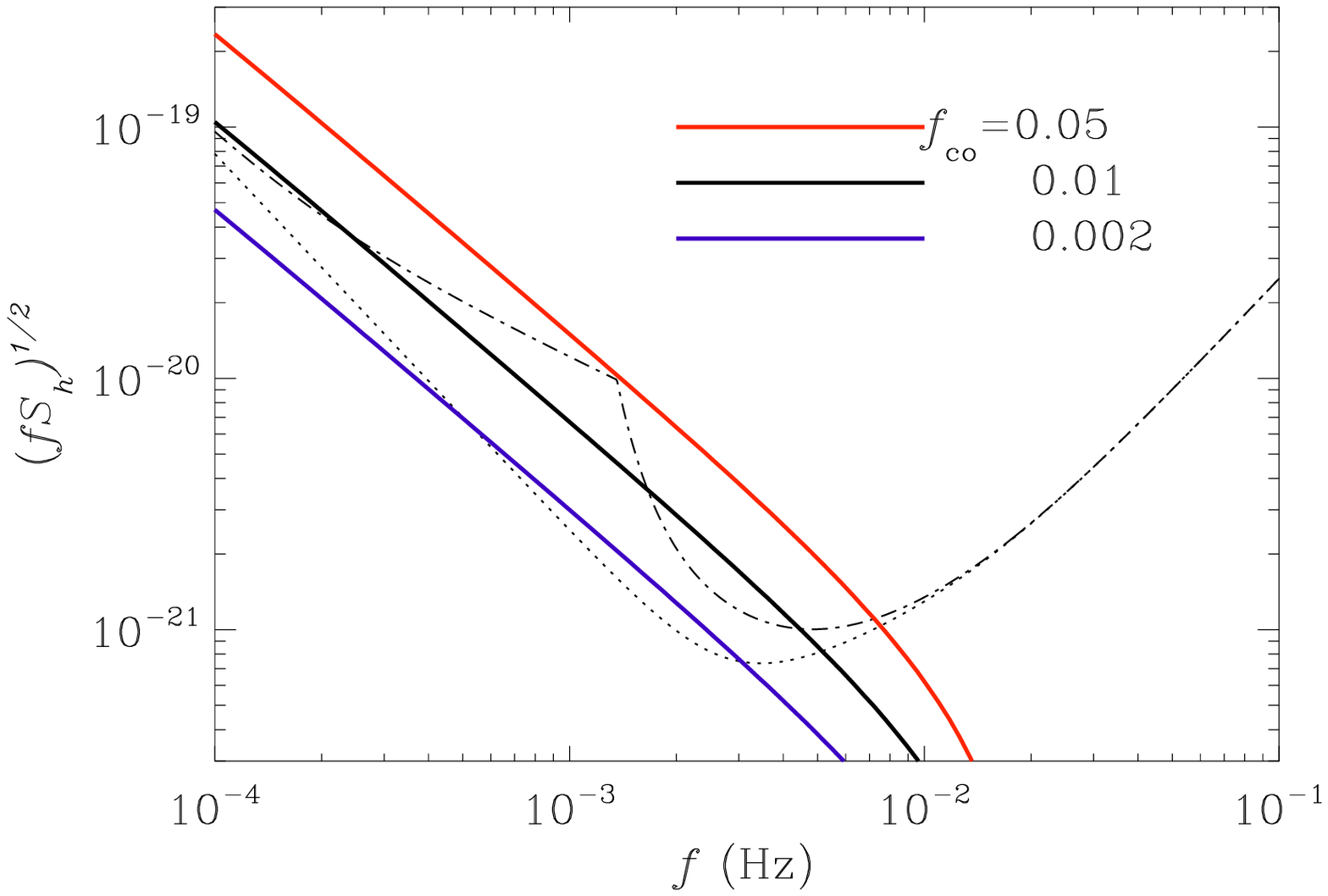}\\
\includegraphics[width=0.45\textwidth,clip=true]{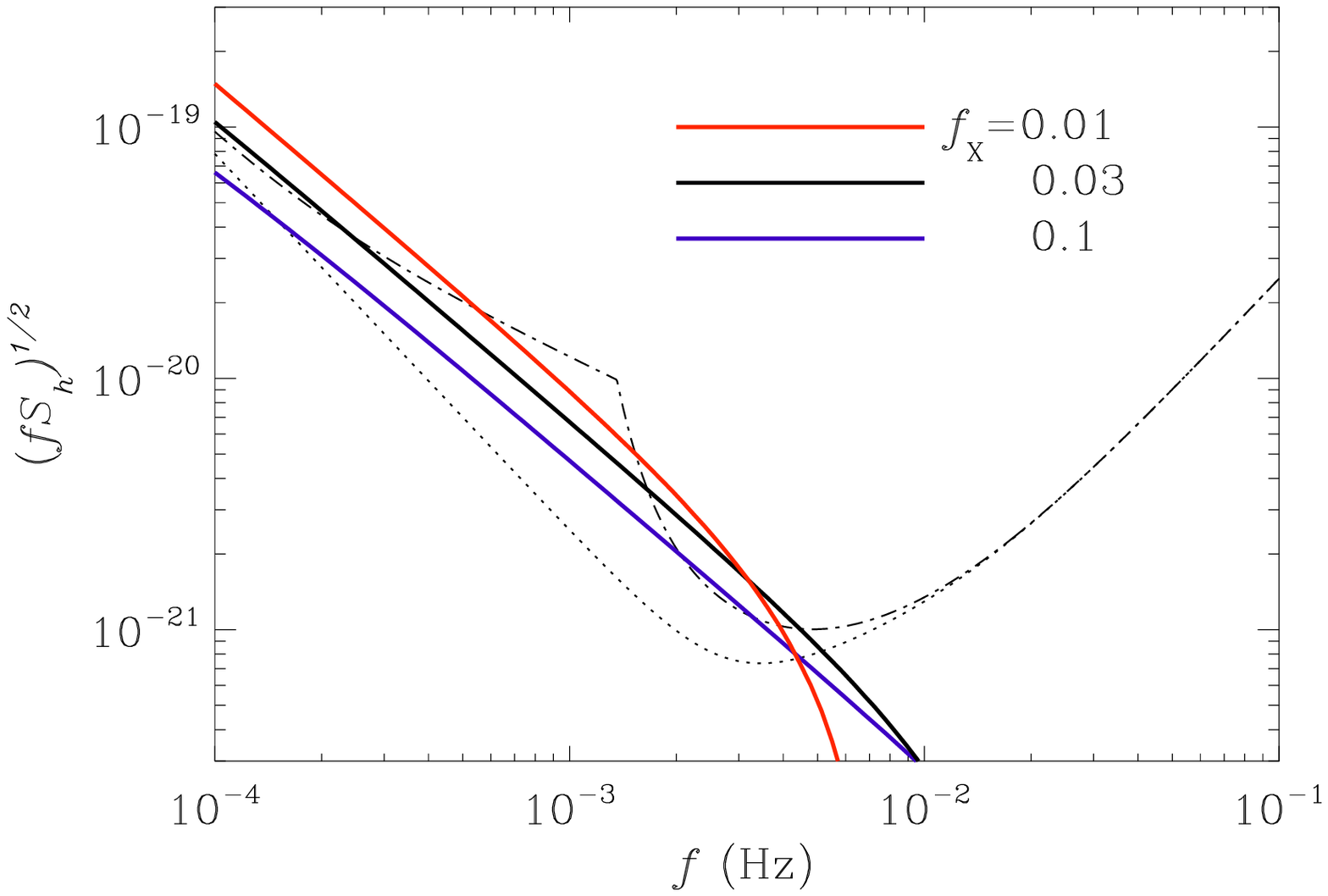}
\includegraphics[width=0.45\textwidth,clip=true]{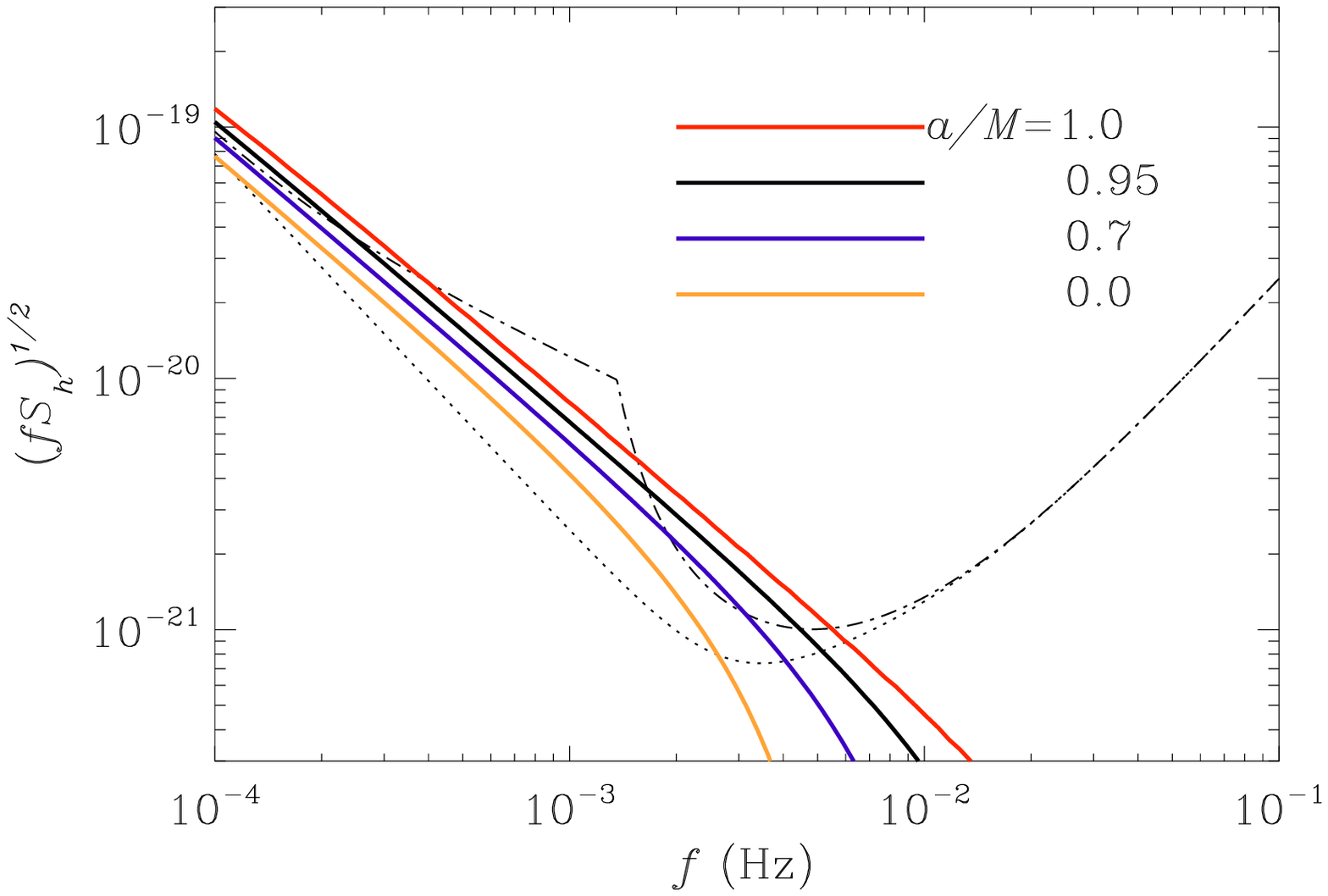}
\caption{\label{results1} Time-averaged GW spectra for the baseline
  model parameters listed in Table \ref{glossary} ({\it solid black
  curves}), and the effects of changing the various parameters:
  $f_{\rm Edd}$ ({\it top left panel}), $f_{\rm co}$ (top right), $f_{\rm X}$
  (bottom left), and $a/M$ (bottom right). The {\it dot-dashed} and {\it
  dotted} lines are the sky-averaged LISA noise curves with and
  without the contributions from galactic binaries, respectively.}  
\end{figure}

In this section we show the impact of each of the model
parameters on the predicted GW spectrum. In this approach, all
parameters are taken as constant in time and independent of the
other parameters and the AGN masses.
In Fig.~\ref{results1}, we show the time-averaged GW spectrum
calculated from Eq.~(\ref{Omega3}), for the fiducial model
parameters listed in Table \ref{glossary} (solid black curves), along with
the effects of varying the individual parameters around these baseline
values. 

In the top left panel of Fig.~\ref{results1} we show the dependence on $f_{\rm Edd}$ relative
to the fiducial model. For smaller values of $f_{\rm Edd}$, the
observed luminosity function will imply a SMBH mass distribution
function shifted towards higher masses and thus lower frequencies
(blue curve). Conversely, higher values of $f_{\rm Edd}$ require
smaller black hole masses and thus higher frequencies (red curve).

The top right panel of Fig.~\ref{results1} shows the dependence on $f_{\rm co}$, the fraction
of accreted matter in the form of compact objects. The linear
dependence of $\Omega_{\rm gw}(f)$ on $f_{\rm co}$ is reasonable: more
compact objects give more signal, which is also clear from 
Eq.~(\ref{Omega3}). Similarly, the predicted GW power increases when the
fraction $f_{\rm X}$ {\it decreases}, as seen in the bottom left panel of 
Fig.~\ref{results1}. The largest GW signal in this scenario
is achieved for minimal $f_{\rm X}$. This corresponds to only a small
fraction of the total accretion activity in the Universe being
observable in hard X-rays. Thus we infer that the bolometric
luminosities are actually much higher, which in turn implies more
matter being accreted and thus more compact objects for a fixed
$f_{\rm co}$. The fraction $f_{\rm X}$ also determines the AGN mass cutoff
through Eq.~(\ref{Lx}), since (fixing all other parameters) a
smaller $f_{\rm X}$ implies a larger Eddington luminosity, and thus
higher mass and lower frequency. 

Motivated by Refs.~\cite{gammie04,mckinney04}, we have used a large
(yet not maximal) spin parameter of $a/M=0.95$ for most of the
calculations here. The bottom right panel of Fig.~\ref{results1} 
shows the fiducial model along with the corresponding signal for $a/M=0,$ $0.7$, and $1.0$. 
Despite the significant difference between the signals from individual
inspiral events (see Fig.~\ref{single_inspiral}), the difference in total,
time-averaged power is rather smaller. This is because rapidly
spinning black holes are more efficient (high
$\eta_{\rm em}$ and $\eta_{\rm gw}$), so for a given luminosity, there
is less gas being accreted and therefore fewer embedded compact
objects, and less GW power. Similar to the
results in the bottom left panel of Fig.~\ref{results1}, varying $a/M$ also changes the
inferred SMBH mass for a given luminosity, in turn changing the cutoff
frequency. 
 
We should mention that the GW signal appears to depend insignificantly
on the maximal redshift $z_{\rm max}$ 
for $z_{\rm max}\gtrsim3$, where the luminosity function becomes
more and more uncertain.
Another parameter that we have explored is the low-luminosity cutoff
of the luminosity function Eq.~(\ref{lumin1}). By extending the cutoff
to lower luminosities, we include lower-mass AGN, and thus higher
frequencies. After trying $L_{\rm min}=10^{40.5}$ ergs s$^{-1}$ and
$L_{\rm min}=10^{39.5}$ ergs s$^{-1}$ we found that, as expected, the lower
luminosities give more signal at higher frequencies, but not
significantly within the LISA band. However, this may be an important
factor when designing a GW observatory with more sensitivity in the
$\sim 0.1-10$ Hz band. On the other hand, the smaller accretion disks
and shorter time scales of low-mass AGN may make them insignificant
sources of disk-embedded compact objects, and thus the somewhat
artificial cutoff of $L_{\rm min}=10^{41.5}$ ergs s$^{-1}$ may
actually be physically justified.

\subsection{Sensitivity to varying model parameters}

\begin{figure}[ht]
\includegraphics[width=0.45\textwidth,clip=true]{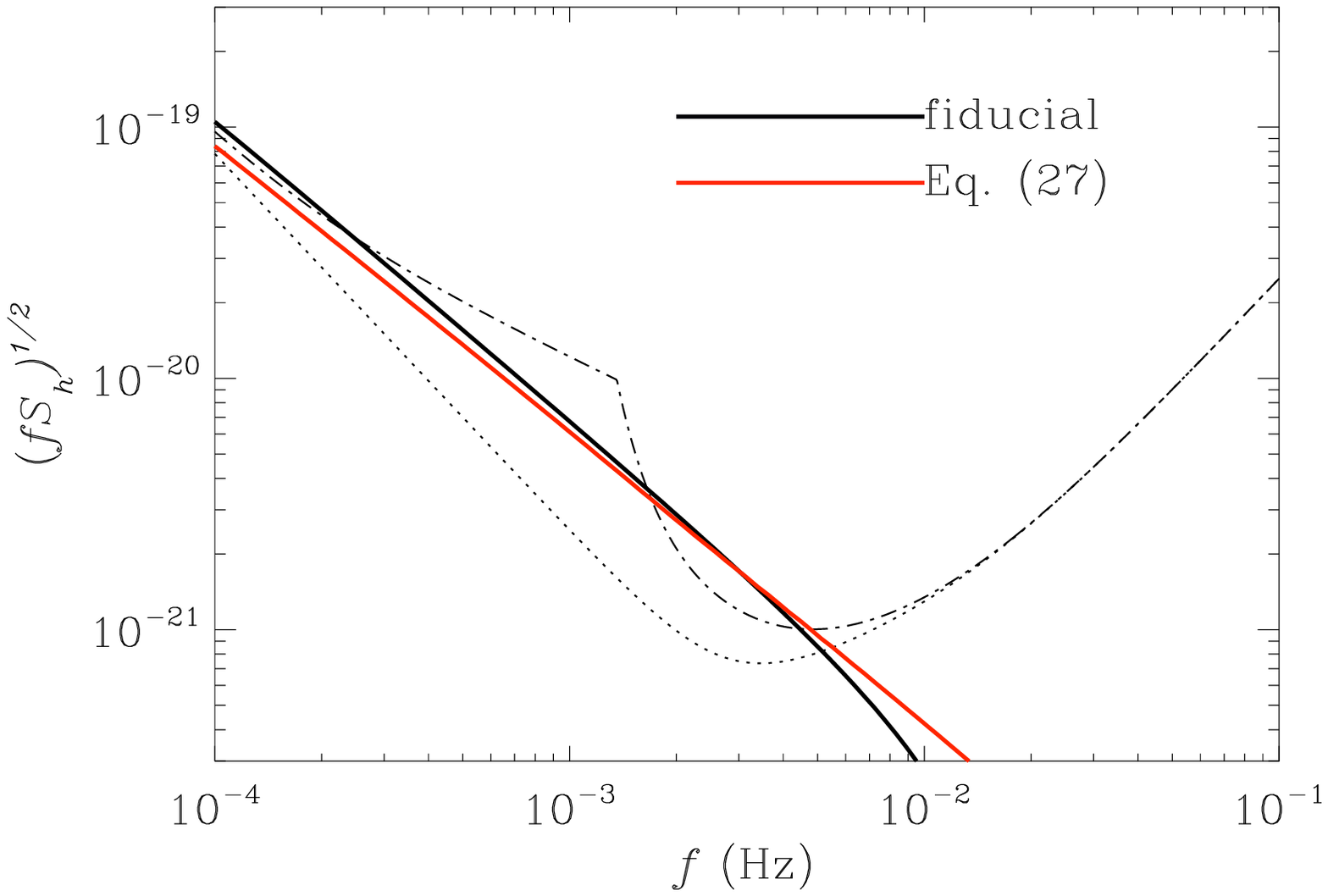}
\includegraphics[width=0.45\textwidth,clip=true]{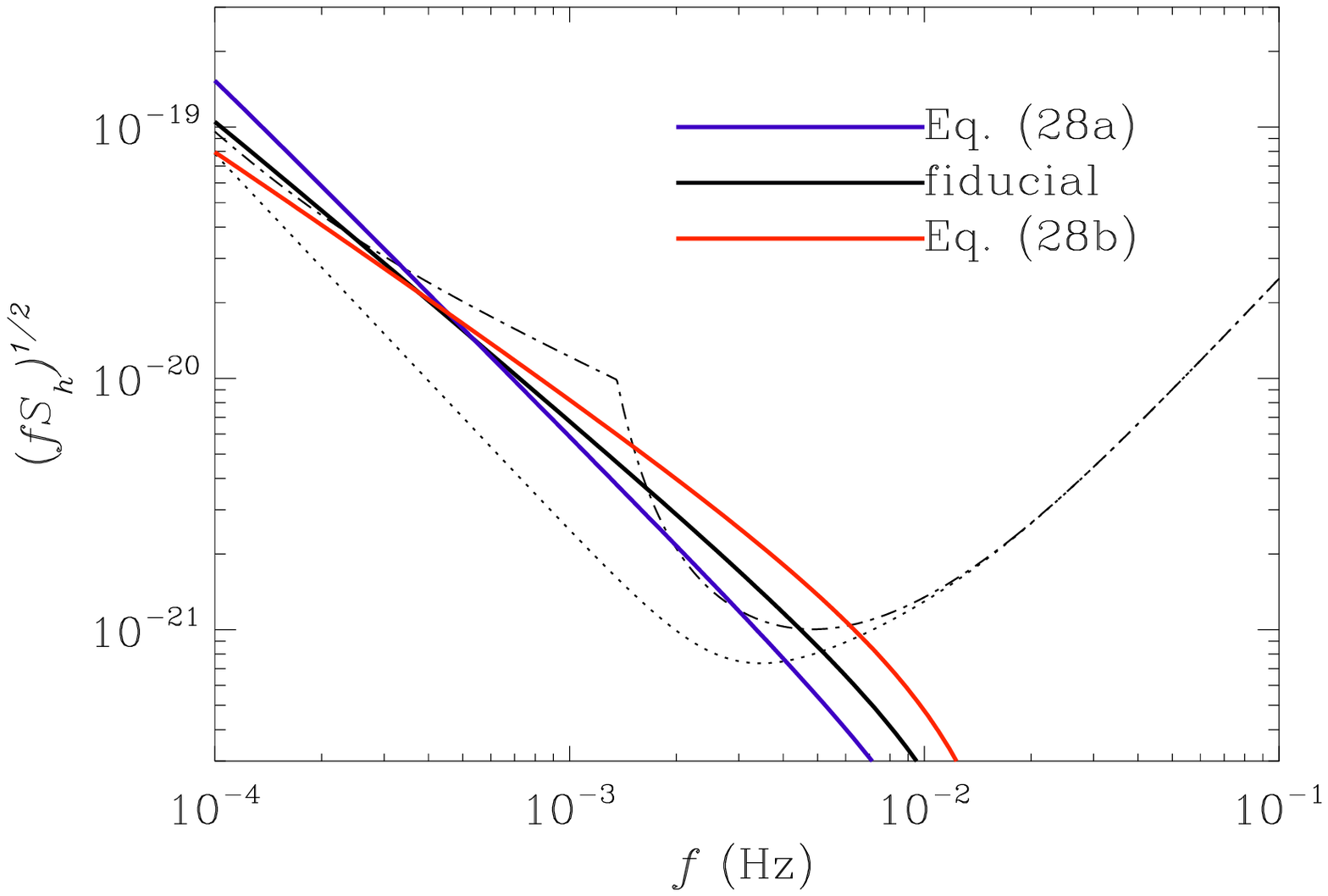}\\
\caption{\label{results2} Time-averaged GW spectra for the baseline
model parameters listed in Table \ref{glossary} (solid black curves), along
with models where
$f_{\rm Edd}$ varies according to Eq.~(\ref{var_Edd})
(left panel, red curve) or 
the fraction $f_{\rm co}$ of accreted matter in form of compact
objects varies according to Eq.~(\ref{var_co}) (right panel,
red and blue curves).}
\end{figure}

Up to now we have assumed that the fiducial values of the parameters
in Table \ref{glossary} are constants independent of SMBH mass and/or
redshift. Particular parameters are, however, likely to vary throughout
evolution. There is evidence, for example, that the Eddington ratio
$f_{\rm Edd}$ was higher in the past than it is now, see, e.g.,
Refs.~\cite{Miller:2006tw,Netzer:2006rg}. In order to assess the
possible influence of such evolution, following Ref.~\cite{Netzer:2006rg}
who studied an AGN sample from the Sloan Digital Sky Survey,
we model the dependence of the average $f_{\rm Edd}$ on the SMBH mass and 
redshift as follows (while restricting $0.01\le f_{\rm Edd}\le 1.0$):
\beq
f_{\rm Edd}(M,z)=0.1\left(\frac{z}{0.1}\right)^{\gamma(M)}
\,\left(\frac{M}{10^7\,M_\odot}\right)^{-0.8}\,,\label{var_Edd}
\eeq
where $\gamma(M)$ is given by Eq.~(4), Table~1, and Fig.~4
in Ref.~\cite{Netzer:2006rg}. This is, in fact, probably more realistic
than the constant $f_{\rm Edd}\simeq0.1$ approximation, because the
SMBH density resulting from the X-ray luminosity
function at small redshift $z\sim0.01$,
$\Omega_{\rm SMBH}\simeq1.3\times10^{-6}$ is closer to
the one inferred from the velocity dispersions in the AGNs: Note
that $\Omega_{\rm SMBH}\sim\int dL_X (dn/dL_X) M/\rho_c$, where
$M$ is related to $L_X$ via the generalization of Eq.~(\ref{Lx})
to varying $f_{\rm Edd}$, $L_X=f_X f_{\rm Edd}(M,z)L_{\rm Edd}(M)$,
and $M$ is restricted to
$10^5\,M_\odot\lesssim M\lesssim 10^{10}\,M_\odot$.
The time-averaged GW spectrum resulting from Eq.~(\ref{var_Edd})
is shown in the left panel of Fig.~\ref{results2}. Compared to the
fiducital model (black curve), when varying $f_{\rm Edd}$ (red curve),
the slope becomes
smaller, i.e. the signal at low frequency decreases whereas
at high frequencies it increases, consistent with the tendencies
seen in the top left panel of Fig.~\ref{results1}. 
Overall, the spectrum seems relatively robust with respect to
variation of $f_{\rm Edd}$. 

Furthermore, since the larger mass disks seem more susceptible to
fragmentation into smaller mass objects, $f_{\rm co}$ might increase
with $M$, whereas $m$ might decrease with $M$. For large $M$, the
longer accretion times also make it more likely for lower-mass stars
to form and evolve to compact objects before reaching the inner edge
of the disk. On the other hand, the feedback into the disk from star
formation may increase gravitational stability and thus decrease
$f_{\rm co}$. For purpose of illustration we consider each of the
following scalings:
\begin{subequations}\label{var_co}
\begin{eqnarray}
f_{\rm co}(M)&=&10^{-3}\left(\frac{M}{10^5\,M_\odot}\right)^{0.5}
\qquad {\rm and} \\
f_{\rm co}(M)&=&10^{-3}\left(\frac{10^9\,M_\odot}{M}\right)^{0.5},
\end{eqnarray}
\end{subequations}
in all cases requiring $f_{\rm co}\lesssim 0.5$ to maintain a coherent
gas accretion disk in order to produce the necessary X-ray flux.
The resulting time-averaged GW spectrum is shown in the right panel 
of Fig.~\ref{results2}. When $f_{\rm co}$ increases with $M$, the
signal is considerably enhanced at low frequencies (blue curve), while
the opposite occurs when $f_{\rm co}$ decreases with $M$ (red curve).

\section{Event rates, duty cycle and confusion noise}
\label{duty_cycle}

The total inspiral rate as seen from Earth at frequency $f$ can be
written as (see, e.g.\ Ref.~\cite{phinney01})
\begin{equation}\label{rate}
  \Gamma(f)= \int_{0}^{\infty}dz\,\frac{R(z)}{1+z}\frac{dV}{dz}
  =\frac{f_{\rm co}}{f_{\rm X}\eta_{\rm em}m}\,
  \int_{0}^{\infty}dz \frac{4\pi r^2(z)}{(1+z)H(z)}
  \int_{L_{\rm min}}^{L_{\rm max}(f_z)} dL_{\rm X}\frac{dn}{d\ln L_{\rm X}},
\end{equation}
where $dV/dz=4\pi r^2(z)/H(z)$ is the fractional volume element,
the Hubble rate at redshift $z$ is given by Eq.~(\ref{cosmo}), and
$r(z)$ is the comoving coordinate, $dr=(1+z)dt$. In the second expression
of Eq.~(\ref{rate}) we have used Eq.~(\ref{R_z}) with
$E_{\rm gw}=\eta_{\rm gw} m$ and $L_{\rm max}(f)$ is the maximum
luminosity for which the associated SMBH mass emits to frequencies
up to $f$. Note that for a fixed compact object fraction
$f_{\rm co}$,
the rate Eq.~(\ref{rate}) is inversely proportional to the typical
mass $m$ of the inspiraling compact object. 

The {\it duty cycle} ${\cal D}(f)$ at a given frequency is the average number of sources
contributing at any given time. This can be estimated by multiplying the
integrand of Eq.~(\ref{rate}) with the time scale over which a source
radiates around the frequency $f$ with a coherent phase development.
In the case of adiabatic, circular inspirals, the system emits GWs at a
well-defined frequency $f(t)$ which evolves monotonically in time.
In the local rest frame (i.e.\ ignoring cosmological redshifts) we can
estimate the coherence time as~\cite{peters64}:
\begin{equation}\label{T_coh}
  t_{\rm coh}(f)\equiv \frac{f}{df/dt} \simeq
  \frac{5}{144\,M^{2/3}\,m\,(\pi f)^{8/3}}
  \simeq 3.5\left(\frac{10^6\,M_\odot}{M}\right)^{2/3}
  \left(\frac{10^2\,M_\odot}{m}\right)
  \left(\frac{10^{-3}\,{\rm Hz}}{f}\right)^{8/3}\,{\rm yr}.
\end{equation}
The duty cycle as observed at Earth will thus be estimated by
multiplying the integrand of Eq.~(\ref{rate}) with $(1+z)t_{\rm
  coh}(f_z)$:
\begin{equation}
  {\cal D}(f) =\frac{f_{\rm co}}{f_{\rm X}\eta_{\rm em}m}
  \int_{0}^{\infty} \frac{4\pi r^2(z)}{H(z)}dz\,
  \int dL_{\rm X}\frac{dn}{d\ln L_{\rm X}}t_{\rm coh}(f_z)\,. 
\label{duty}
\end{equation}
Note that it is proportional to $f_{\rm co}/(f_{\rm X}\eta_{\rm
  em}m^2)$, with the extra power of $m$ coming from the dependence of
  $t_{\rm coh}$ in Eq.~(\ref{T_coh}). Smaller $m$ means more compact
objects, and also slower inspiral rates, thus each source spends more
time around a given $f$.

\begin{figure}[ht]
\includegraphics[width=0.55\textwidth,clip=true]{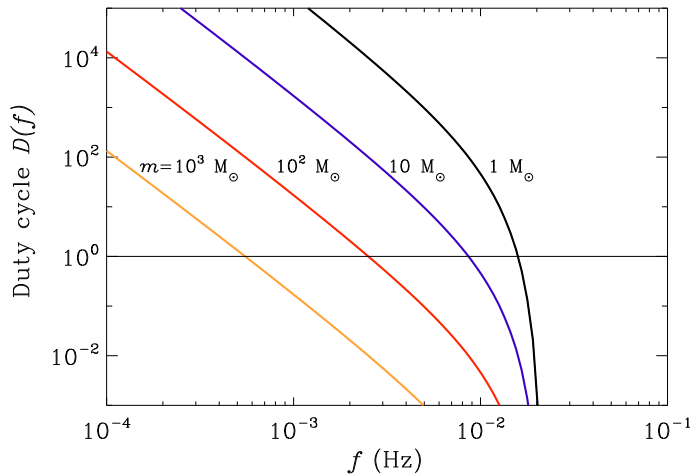}
\caption{\label{D_f} Duty cycle ${\cal D}(f)$ for the nominal
  parameter values and a range of compact object masses $m$. When
  ${\cal D}(f)\lesssim 1$, the inspiral signals should be individually
  resolvable. The high frequency cut-off is due to the somewhat arbitrary
  cut-off of the X-ray luminosity function at low luminosities
  which depends on the experimental sensitivity.}
\end{figure}

In Fig.~\ref{D_f} we show the duty cycle for the fiducial
model parameters and a range of compact object masses $m$. The
cutoff around 20 mHz is due to the somewhat artificial low-end
cutoff in the luminosity function, corresponding to a minimum value
for $M$ and thus maximum attainable frequency.
When the duty cycle is $\lesssim1$, a detector will see a
non-gaussian, non-continuous signal which rather has a ``popcorn''
character~\cite{drasco02}. Provided the signal to noise ratio
(SNR) is sufficiently large, the detector will then be able to resolve
individual
events of duration $t_{\rm coh}(f)$, occurring at a rate
$\Gamma(f)<t_{\rm coh}^{-1}(f)$. 

When the duty cycle is $\geq \, 1$, several sources are present at any
given time at a frequency $f$ in a bandwidth $\Delta f \sim
f$. However, the number of sources in the smallest resolution
frequency-bin, which corresponds to $1/T_{\rm obs} \sim 10^{-8}$ Hz,  
is not larger than one. Thus, our background is in principle
subtractable if the SNR is sufficiently high.

\begin{figure}[ht]
\includegraphics[width=0.45\textwidth,clip=true]{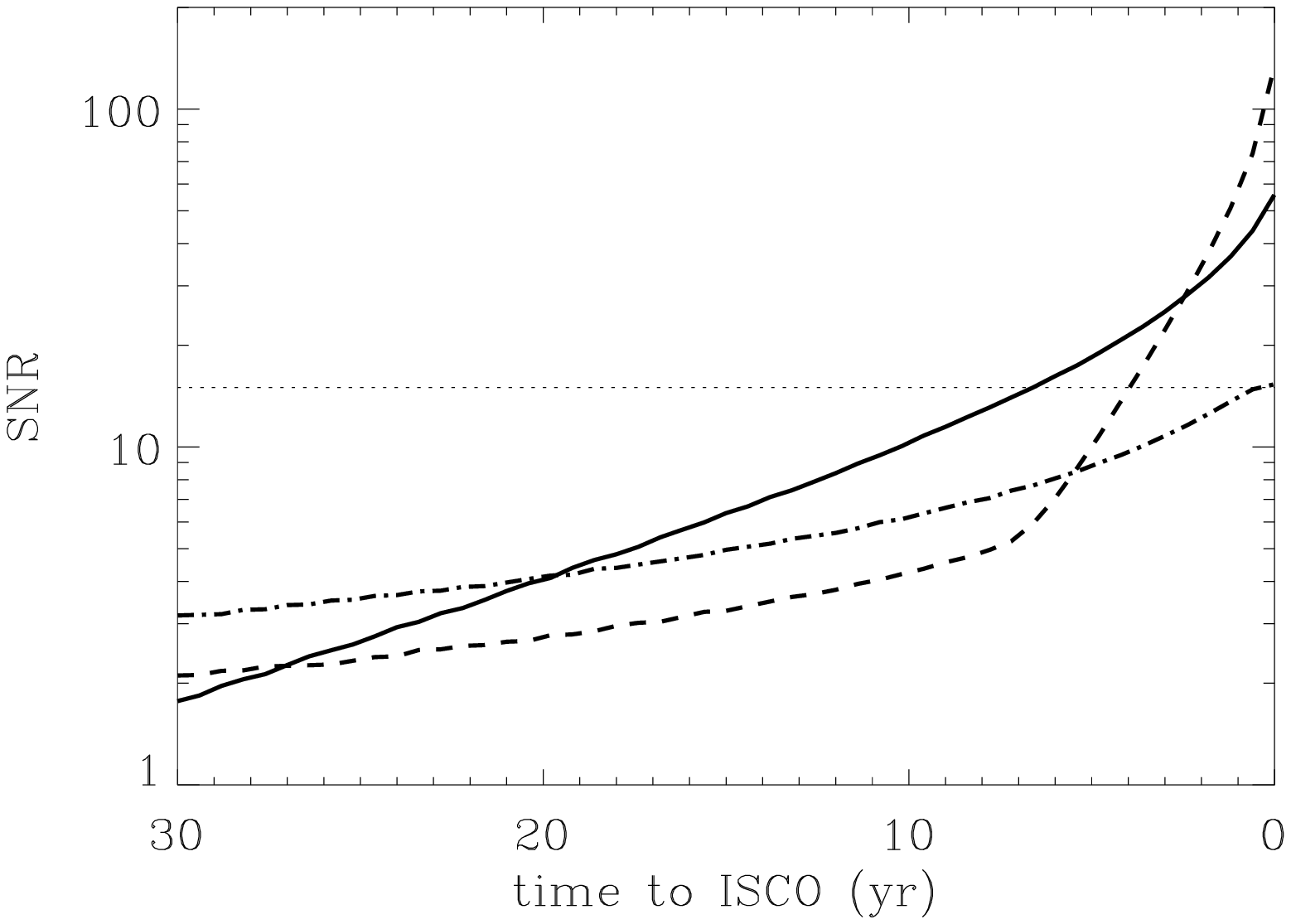}
\includegraphics[width=0.45\textwidth,clip=true]{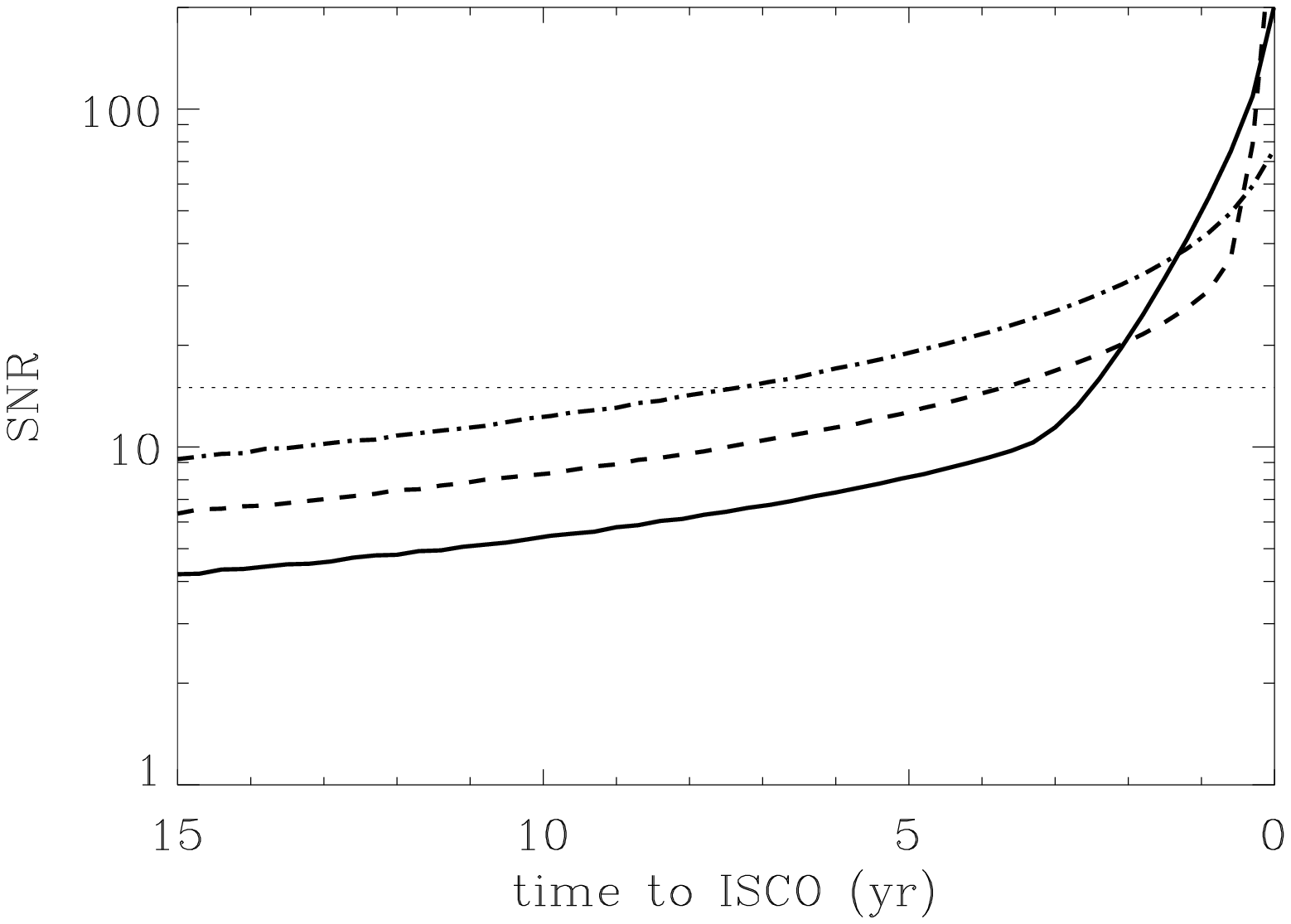}
\caption{\label{snr_fig} Signal to noise ratio for the inspiral of a
  compact object of masses ({\it left panel}) $10 M_\odot$ and ({\it
  right panel}) $100 M_\odot$
  onto a SMBH of mass $M=10^5\,M_\odot$ (solid line),
  $M=10^6\,M_\odot$ (dashed line), and $M=10^7\,M_\odot$ (dot-dashed
  line) at 1 Gpc distance, as a function of the time left to the ISCO
  at the end of the observation time $T_{\rm obs}=3\,$yr. Also shown
  is the threshold of SNR=15 (dotted line).}
\end{figure}

Let us estimate, as done in Ref.~\cite{barack04}, the unsubtractable
portion which will constitute the {\it confusion} noise $S^{\rm
  conf}(f)$ [or $d\rho_{\rm gw}^{\rm conf}(f)/df$ related to $S_h^{\rm
    conf}(f)$ by the analogue of Eq.~(\ref{Srho})].
For each inspiral source observed over the LISA mission lifetime, we
evaluate the SNR defined by 
\begin{equation}\label{snr} 
{\rm SNR}^2 = \int_{f_{\rm min}}^{f_{\rm max}} d \ln f\,
\frac{h^2_c(f)}{f\,S_{\rm n}}\,,
\end{equation}
where $h_c(f)$ is the characteristic strain defined in Eq.~
(\ref{hc}) and $S_{\rm n}$ is given by Eq.~(\ref{Shtot}). This
integral is performed over the frequency range $f_{\rm min}\leq
f\leq f_{\rm max}$ through which the system sweeps during the mission
lifetime $T_{\rm obs}$. While in principle the inspiraling object may
reach $f_{\rm ISCO}$ and then plunge sometime in the middle of the
LISA mission, for concreteness we only consider systems that are
active throughout $T_{\rm obs}$. 
In Fig.~\ref{snr_fig} we show SNRs for a few typical cases as a
function of the time $t$ remaining to plunge (as of the end of the
observation).
We then insert a step-function factor $\Theta\left({\rm
  SNR_{thr}}-{\rm SNR}\right)$ in the integrand of Eq.~(\ref{Omega3})
to obtain $d\rho_{\rm gw}^{\rm conf}(f)/df$. 
This assures that only events with SNR too small to be subtracted out contribute
to this confusion noise.

We note in passing that, strictly speaking, unresolvable events contribute
to the $S_{\rm n}$ noise and thus the SNR in Eq.~(\ref{snr}) should be obtained
by an iterative procedure in which $S_{\rm n}$ includes events that could
not be subtracted out. This becomes important when the merger
noise of events whose SNR is smaller than the threshold SNR is comparable or
larger than the instrumental noise. In this case we will slightly underestimate
the fraction of the true unsubtractable background.

\begin{figure}[ht]
\includegraphics[width=0.45\textwidth,clip=true]{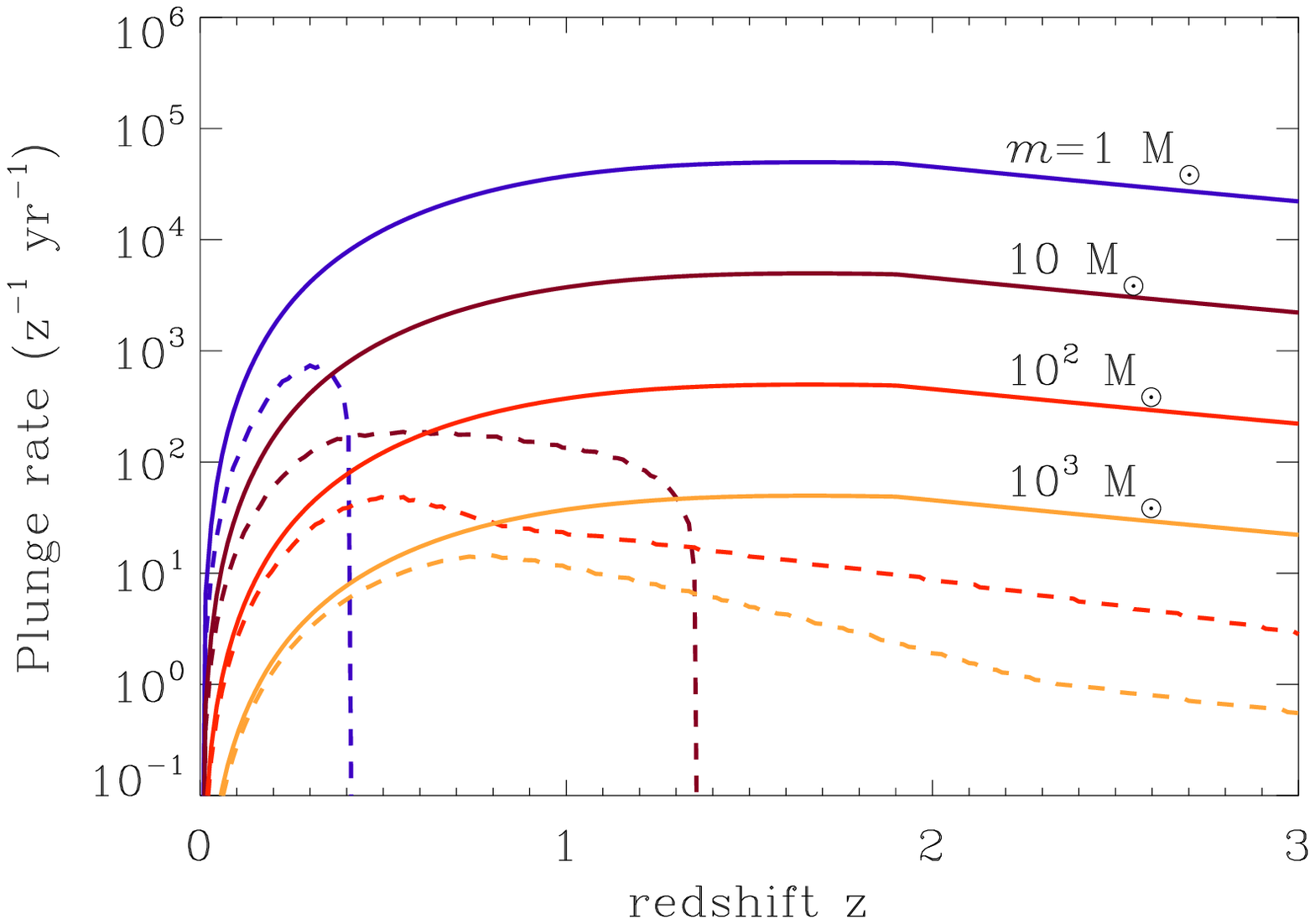}
\includegraphics[width=0.45\textwidth,clip=true]{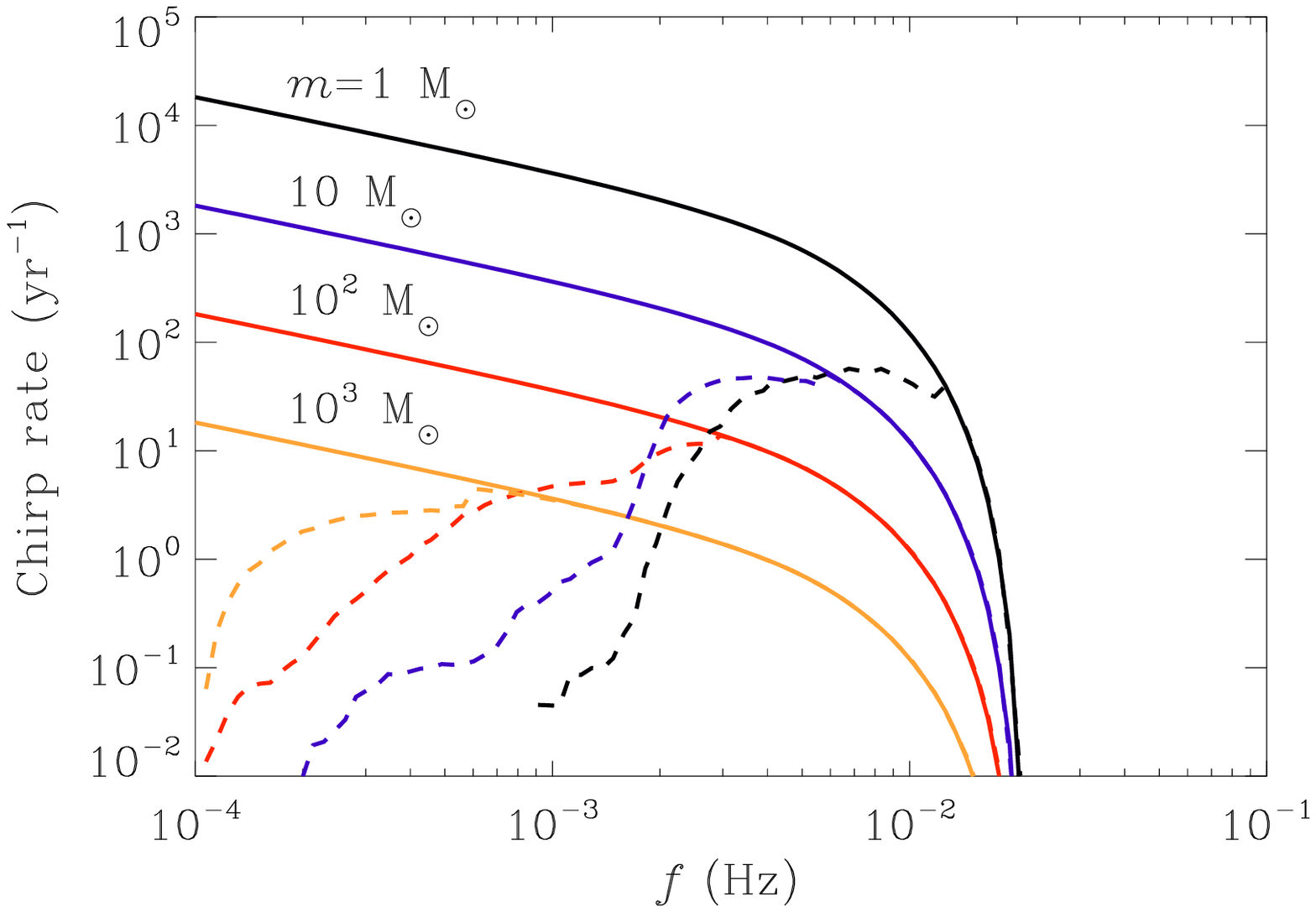}
\caption{\label{inspiral_rate} In the left panel we show 
  the inspiral rate $\Gamma(z)$ as a function of redshift for $m=1-10^3 M_\odot$.
  The solid curves represent the total plunging rate in the
  Universe {\it irrespective of frequency}, while the dashed curves
  represent the resolvable fraction with maximal SNR $\ge 15$
  In the right panel we show the total inspiral chirp
  rate $\Gamma(f)$ of Eq.~(\ref{rate}) for the same range of masses,
  again distinguishing between the total rate in the Univers (solid
  curve) and the resolvable portion (dashed curves).}
\end{figure}

With the SNR calculation in hand, we can also estimate the inspiral
rate above a given threshold. In the left panel of
Fig.~\ref{inspiral_rate}, we show
both the total inspiral rate in the Universe (solid curves), as well
as the rate of detectable signals for SNR $\ge 15$ (dashed curves),
for $m=1-10^3 M_\odot$. The total rate, which here can be thought of
as the rate of compact objects reaching the ISCO, is given at each
redshift as 
\begin{equation}
\Gamma(z) = \frac{R(z)}{1+z} \frac{dV}{dz}.
\end{equation}
We have also defined [see Eq.~(\ref{rate})] an inspiral rate as a
function of frequency 
$\Gamma(f)$, which is a measure of the number of chirping systems that
passes through a frequency $f$ per unit time.
To distinguish these
rates, we will refer to $\Gamma(z)$ as the ``plunge rate'' and
$\Gamma(f)$ as the ``chirp rate.'' 

While the plunge rate is proportional to $m^{-1}$, the SNR increases with
$m$, so the smaller mass inspirals are only resolvable out to smaller
distances. Also, at a given redshift, the integrated SNR generally
decreases with increasing $M$ as the lower $f_{\rm ISCO}$ resides in
the region of higher instrumental noise. By integrating the area under
the dashed curves in Fig.~\ref{inspiral_rate}, we find
the resolvable event rates to be $\sim 150 {\rm yr}^{-1}$ for
$m=1-10 M_\odot$ and $\sim 50 {\rm yr}^{-1}$ for $m=100
M_\odot$. Inspirals with $m=10^3 M_\odot$, while observable out to
high redshifts, would be relatively rare, with rates $\sim 15 {\rm
 yr}^{-1}$. From Eq.~(\ref{rate}), we see that these plunge rates are
proportional to $f_{\rm co}$, so could potentially be used to
constrain that poorly-known parameter.

In the right panel of Fig.~\ref{inspiral_rate} we plot the chirp rate
$\Gamma(f)$ for the same range of compact object masses. At low
frequencies, the chirp
rate funtion is nearly flat, due to a steady-state ``flux
conservation'' as each inspiraling object enters a frequency bin,
another will leave it. Then, at higher frequencies, systems begin to
drop out altogether as they reach the plunge frequency for a given
mass $M$, up to the final cut-off frequency due to the lower limit end
of the SMBH mass function (compare with the duty cycle ${\cal D}(f)$
in Fig.~\ref{D_f}). 

Note that these event rates are, within orders of magnitude,
consistent with independent estimates of EMRIs
based on loss cone calculations. Such estimates give inspiral rates of
``captured'' compact
objects per galaxy ranging from $\sim10^{-7}$~\cite{Hopman:2006xn} to
$\sim10^{-5}\,{\rm yr}^{-1}$~\cite{Holley-Bockelmann:2006jp}.
With a few billion galaxies out to $z\sim 1$, this
corresponds to rates between 
$\sim100$ and $\sim10^4\,{\rm yr}^{-1}$.

\begin{figure}[ht]
\includegraphics[width=0.55\textwidth,clip=true]{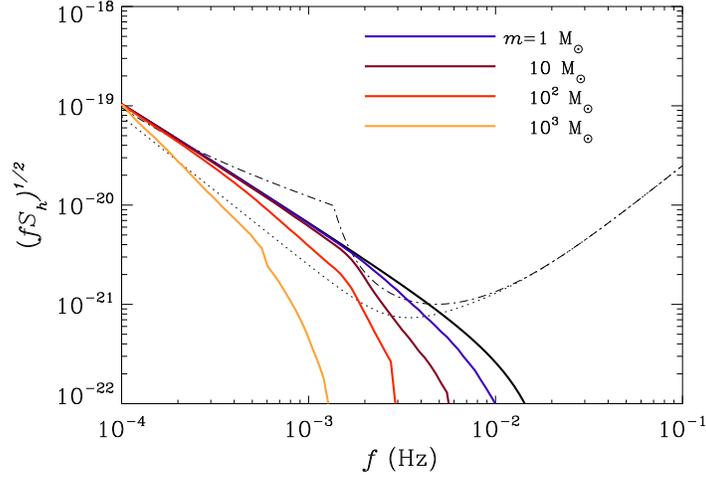}
\caption{\label{unresolvable} Unresolvable GW spectra for a range of
  compact object masses $m$, after
  subtracting out individual events with integrated SNR above a
  threshold of 15. The solid black curve corresponds to the total
  time-averaged spectrum, including the resolvable portion.}
\end{figure}

In Fig.~\ref{unresolvable} we show the unresolvable portion of the
GW power spectrum for the fiducial model and a range of $m$,
subtracting out all contributions from systems with SNR>15. Clearly,
for larger values of $m$, the threshold SNR is reached at an earlier
time and thus lower frequency in the inspiral, reducing the confusion
noise at higher frequencies. We notice that even the subtracted events
can increase the total noise 
(\ref{Shtot}) by an amount proportional to $S_n$~\cite{Cutler}, because 
information used to subtract the events cannot be used to detect 
other events. However, in our case, we have found that this increase 
in negligible. 

From the dependence ${\cal D}(f)\propto f_{\rm co}/m^2$ and from
Fig.~\ref{D_f}, we can deduce that for fixed $f_{\rm Edd}$ and
$\eta_{\rm em}$ and $m\lesssim10^2\,(f_{\rm co}/0.01)^{1/2}\,M_\odot$,
the duty cycle is larger than unity and, therefore, the background can
basically be treated as gaussian at all frequencies where a detectable
signal is
predicted. For $m\gtrsim10^2\,(f_{\rm co}/0.01)^{1/2}\,M_\odot$, the
frequency where the duty cycle crosses unity is in a range where
${\cal D}(f)$
roughly scales as $f^{-8/3}$ (see Fig.~\ref{D_f}), because the inspiral
rate $\Gamma(f)$ varies much more weakly (see Fig.~\ref{inspiral_rate}),
and the coherence time,
Eq.~(\ref{T_coh}), $t_{\rm coh}\propto f^{-8/3}$. As a result, the GW
background becomes gaussian when
\begin{equation}\label{f_gaus}
  f\lesssim f_{\rm gauss} \simeq 2\times10^{-3}\,
  \left(\frac{f_{\rm co}}{0.01}\right)^{3/8}\,
  \,\left(\frac{10^2\,M_\odot}{m}\right)^{3/4}\,{\rm Hz}
\end{equation}

Note from Fig.~\ref{unresolvable} that the frequency below which the
background becomes unsubtractable, and thus contributes to confusion
noise, is lower 
by a factor $\sim$5--10 compared to the frequency Eq.~(\ref{f_gaus}) 
below which the noise becomes gaussian because the duty factor 
becomes larger than unity. Figure \ref{unresolvable}
also shows that the compact object mass above which inspirals tend
to become resolvable, $m\gtrsim10\,M_\odot$, is comparable to the
compact object mass above which the signal consists of individual
events (i.e.\ the duty cycle is smaller than unity).
In other words, there is indeed a range of frequencies and compact
object masses giving a quasi-continuous signal which may still be
resolved into individual events given an improved detector. However,
when the duty cycle is much greater than unity, even a perfect
detector will not be able to resolve the individual sources, as is the
case for most of the galactic white dwarf population \cite{cornish01}.

\vspace{0.5cm}

\section{Conclusions}\label{discussion}

Assuming scaling relations between accretion rates, intrinsic
bolometric and X-ray luminosities, and supermassive black hole mass on
the one hand, and rates and masses of compact object inspirals on the
other hand,
one can predict GW spectra from AGN sources. When part
of these scaling relations are constrained observationally, one
obtains amplitudes detectable by LISA if the fraction of material
accreted in the form of compact objects is $\gtrsim1\%$.  The
GW signal depends relatively little on the accretion
and emission parameters, as long as these are within the observational
constraints.  It increases with the fraction of the observed mass
density of supermassive black holes that is due to accretion $(f_{\rm
  Edd})$, and is 
thus largest for accretion dominated growth. 

Our scenarios are well within existing observational upper limits. For example,
for the highest flux shown in Fig.~\ref{results1} one obtains
$\Omega_{\rm gw}(10^{-8}\,{\rm Hz})h_0^2\lesssim10^{-12}\,(f_{\rm
  co}/0.01)$, compared to the msec pulsar timing limit
$\Omega_{\rm gw}(10^{-8}\,{\rm Hz})h_0^2\lesssim10^{-8}$~\cite{pulsartiming}.
{This class of sources is not important for future missions sensitive
around 1 Hz, such as BBO~\cite{bbo}.

If LISA establishes a detection or upper limits on $\Omega_{\rm gw}(f)$,
this may be translated into constraints on the parameters
$f_{\rm co}$, $m$, $\eta_{\rm gw}$, $f_{\rm X}$, $f_{\rm Edd}$, and
$\eta_{\rm em}$ used in the present parametrization. Since $f_{\rm
  X}$, $f_{\rm Edd}$, and $\eta_{\rm em}$ are rather well constrained
by AGN astronomy, and $\eta_{\rm gw}$ is rather well determined by
theory, LISA will mainly constrain $f_{\rm co}$ and $m$. A more
detailed study where these parameters depend on luminosity and/or
different AGN classes may allow for more detailed predictions, but is
beyond the scope of the present exploratory study.

For compact object masses $m\lesssim10^2\,(f_{\rm co}/0.01)^{1/2}\,M_\odot$,
where $f_{\rm co}$ is the fraction of material accreted in form of
these compact objects, the background is gaussian at nearly all
frequencies where a detectable signal is predicted. In contrast, for
$m\gtrsim10^2\,(f_{\rm co}/0.01)^{1/2}\,M_\odot$, the GW background 
is gaussian only at frequencies $f\lesssim2\times10^{-3}\,(f_{\rm
  co}/0.01)^{3/8}\,(10^2\,M_\odot/m)^{3/4}\,$Hz, above which the duty
cylce is smaller than unity. The frequency
below which the GW background becomes confusion noise because the signal
to noise ratio becomes too small to subtract out individual events
is smaller than this gaussian frequency by a factor $\sim$5--10.
At higher frequencies one would observe individual coherent signals of
typical duration
$\simeq0.2\,(10^2\,M_\odot/m)\,(10^{-3}\,{\rm Hz}/f)^{8/3}\,$yr
which occur at a typical rate $\sim 10^2\,(f_{\rm
  co}/0.01)\,(10^2\,M_\odot/m)\,{\rm yr}^{-1}$. 
This is typical for supermassive binary inspiral discussed before in
the literature.

While the time-averaged GW background described in this paper will be
similar to other EMRI backgrounds (e.g.\ \cite{barack04}), the resolvable
waveforms should provide
information that allows us to distinguish between the circular,
equatorial inspirals of disk-embedded objects and the highly inclined,
eccentric orbits predicted for capture inpirals.
When combined with knowledge of the electromagnetic properties of
active galactic nuclei, LISA should be able to constrain primarily the
fraction of accretion in form of compact objects and the mass of these
compact objects, and thus provide fundamental new insights into the
problem of star formation in AGN disks. 

\section*{Acknowledgments} We thank Curt Cutler and Joe Silk for useful discussions.
G.S. thanks the European Network of Theoretical Astroparticle 
Physics ILIAS/N6 under contract number RII3-CT-2004-506222 for partial
financial support and the Physics Department at University of Maryland for 
hospitality during his visit. A.B. acknowledges support from the Alfred Sloan Foundation.

\end{document}